\documentstyle[prl,aps,multicol,epsfig]{revtex}

\newcommand{\equ}{Eq.}
\newcommand{\eqs}{Eqs.}
\newcommand{\fig}{Fig.}
\newcommand{\figs}{Figs.}

\newcommand{\sect}{Sec.}
\newcommand{\app}{App.}
\newcommand{\rem}[1]{}

\newcommand{\bm}[1]{\mbox{\boldmath$#1$\unboldmath}}
\newcommand{\op}[1]{\bm #1}

\newtheorem{platztab}{{Tab.}} % a dummy theorem to count figures!
\rem{

}

\newtheorem{platz}{{Fig.}} % a dummy theorem to count figures!
\newcommand{\FIGo}[3]{\begin{figure}%
#3%
\caption[]{\footnotesize #2}%
\label{#1}%
\end{figure}}
\rem{
\newcommand{\FIGo}[3]{
  \begin{center}
    \begin{minipage}{16.0cm}
    \begin{picture}(16.0,1)
      \put(0,0){\framebox(16.0,1){\ }}
    \end{picture}
      \begin{footnotesize}
        \begin{platz}   \label{#1}
          {\rm #2}
        \end{platz}
      \end{footnotesize}
    \end{minipage}
  \end{center}
  }
}

\rem{
\newcommand{\FIGo}[3]{%
\marginpar{\begin{platz} \label{#1} ~ \end{platz} \vspace*{1.5ex} }
}
}

   \def\bege{\begin{equation}}
   \def\ende{\end{equation}}
   \def\bega{\begin{eqnarray}}
   \def\enda{\end{eqnarray}}
   \def\began{\begin{eqnarray*}}
   \def\endan{\end{eqnarray*}}

\newcommand{\weglassen}[1]{}

\newcommand{\todo}[1]{}

\newcommand{\const}{\mbox{constant}}
\newcommand{\mean}[1]{\langle #1\rangle}
\newcommand{\tauQ}{\tau_{\text Q}}
\newcommand{\tauL}{\tau_{\text L}}
\newcommand{\tauC}{\tau_{\text C}}
\newcommand{\tauK}{\tau_{\text K}}

\newcommand{\bstr}{1}

\begin{document}

\title{The Quantum-Classical Correspondence in Polygonal Billiards}    
\author{Jan Wiersig}
\address{Max-Planck-Institut f\"ur Physik komplexer Systeme,D-01187 Dresden,Germany}
\date{\today}
\maketitle
\begin{abstract} 
We show that wave functions in planar rational polygonal billiards (all angles
rationally related to $\pi$) can be expanded in a basis of quasi-stationary
and spatially regular states. 
Unlike the energy eigenstates, these states are directly related to the
classical invariant surfaces in the semiclassical limit. 
This is illustrated for the barrier billiard.
We expect that these states are also present in integrable billiards with point
scatterers or magnetic flux lines.  
\end{abstract}

\renewcommand{\baselinestretch}{\bstr} \normalsize

\begin{multicols}{2}

\section{Introduction}
\label{sec:intro}
The relation between wave motion in the short-wave\-length regime and the
corresponding ray dynamics is of fundamental importance in quantum mechanics,
electromagnetics and acoustics.   
It has been found in quantum mechanics that the properties of the stationary
solutions of the Schr\"odinger equation, the eigenstates $|E_j\rangle$ of the
Hamiltonian, reflect the degree of order in the classical ray dynamics. 
For example, energy eigenfunctions $\langle x,y|E_j\rangle$ of
classically integrable systems with quasiperiodic motion on invariant tori are
regular, while eigenfunctions of classically chaotic systems with ergodic
motion on energy surfaces are typically irregular.   
For generic systems, it has been conjectured~\cite{Berry77}
(see~\cite{Robnik98} for a review) that in the (semi-)classical limit $\hbar
\to 0$ the averaged Wigner transforms of typical energy eigenfunctions
``condense'' uniformly onto the underlying classical stationary objects in
phase space, which are usually invariant tori, chaotic components or entire
energy surfaces.     
In this article, however, we consider a class of systems with exotic
 classical invariant surfaces for which it is not clear {\it a priori} whether such
 a condensation scenario exists. 
\rem{In this article, however, we demonstrate that for rational polygonal billiards
the quantum-classical correspondence is more complicated: it is not
individual energy eigenstates but certain superpositions of them that are
related to the classical invariant surfaces.}
\rem{for classically long times.
We show that these superpositions have properties similar to energy
eigenstates of integrable systems.}
\rem{
Each such superposition can be regarded as a coherent wave packet localized
around an invariant surface in phase space. The wave packet
varies slowly in time, but it remains in the neighborhood of the
invariant surface for classically long times. 
We show that the wave packets have properties similar to energy
eigenstates of integrable systems.} 
  
The classical free motion inside a planar domain ${\cal Q}$ with elastic
reflection at the boundary has a constant of motion, Hamilton's function $H =
p_x^2+p_y^2+V(x,y)$. The particle's mass is $1/2$, and $V(x,y) = 0$ if $(x,y)
\in {\cal Q}$ and $\infty$ otherwise.  
A billiard with polygonal boundary has a second constant of motion
$K(p_x,p_y)$ if all angles $\alpha_j$ between sides are rationally related to
$\pi$, i.e. $\alpha_j = m_j\pi/n_j$, where $m_j, n_j > 0$ are relatively prime
integers~\cite{Hobson75,ZemlyakovKatok76}.  
However, this does not imply integrability in the sense of
Liouville-Arnol$'$d~\cite{Arnold78}. The Poisson bracket
$\{H,K\}$ vanishes identically only if $m_j = 1$ for all $j=1,2,\ldots$, so only
rectangles, equilateral triangles, $\pi/2, \pi/4, \pi/4$-triangles, and
$\pi/2, \pi/3, \pi/6$-triangles are integrable.  
Critical corners with $m_j > 1$ destroy the integrability in a singular way:
$\{H,K\}$ is zero everywhere in phase space except at a measure-zero set
corresponding to the critical corners;     
the phase space is foliated by two-dimensional invariant surfaces $H,K =
\const$ as in integrable systems, but they do not have the topology of
tori. The motion is not quasiperiodic and characterized as
pseudointegrable~\cite{RichensBerry81}.      

The quantum mechanical free wave motion with Di\-richlet boundary conditions on
a rational polygon barely reflects the classical
pseudointegrability if only individual eigenfunctions of the Hamiltonian
$\op{H} = H(\op{x},\op{y},\op{p_x},\op{p_y})$ are considered.  
A typical eigenfunction is hard to distinguish from eigenfunctions in 
classically chaotic 
systems~\cite{BiswasJain90,BellomoUzer95,VUF95,TomiyaYoshinaga96}.
It is in general not an eigenfunction of the second operator $\op{K} =
K(\op{p_x},\op{p_y})$ and has nonzero uncertainty $\Delta K =
(\mean{\op{K}^2}-\mean{\op{K}}^2)^{1/2}$, since the commutator
$[\op{H},\op{K}]$ does not vanish.  
Only a subtle signature of pseudointegrability seems to be contained in
individual energy eigenfunctions, namely in the distribution of zeroes of the
associated Husimi function~\cite{BS99}.  
The closeness to the quantum mechanics of chaotic systems on the one side
and to the classical dynamics of integrable systems on the other
side is a first hint of an unusual quantum-classical correspondence.

A further indication pointing in this direction comes from an analogy with the
metal-insulator transition of the Anderson model in three dimensions. 
Both kinds of systems have energy level statistics close to the semi-Poisson
distribution~\cite{BGS99}. 
In the Anderson model, at the transition point between extended states with
Wigner statistics and localized states with Poisson distributed energy levels,
this intermediate statistics has its origin in the multi-fractal character of
the wave functions, which are neither extended nor localized~\cite{SG91}.   
Analogously, a typical energy state in a rational polygon is expected
to be a fractal in momentum space~\cite{BGS99}, localized around an energy
surface but neither extended nor localized within this surface; 
condensation onto a lower-dimensional invariant surface is ruled out.    

As a final indication we will present a numerical study on a particular
system, the barrier billiard. 
Instead of Wigner transforms in four-dimensional phase space, we study
distributions in the two-dimensional $(H,K)$-space, where each point
represents a classical invariant surface. In this context, we define
condensation as $\Delta H/\mean{\op{H}} \to 0$ and $\Delta K/\mean{\op{K}} \to
0$ in the semiclassical limit.    
This ensures that mean values of the quantum operators in highly excited
states can be interpreted as well-defined values of the 
classical constants of motion.  
Our central issue is to demonstrate that even though the energy states
presumably do not show such a condensation there is an alternative basis of
states which do so up to classically long times.  

The paper is organized as follows.
In \sect\ref{sec:basis} we construct the basis for general rational
polygons. Its time evolution is discussed in \sect~\ref{sec:time}. 
In \sect~\ref{sec:barrier} we illustrate our statements for the barrier
billiard. 
We conclude with a brief summary in \sect~\ref{sec:con} and an Appendix with
details on the numerical computations.

\section{Construction of the basis}
\label{sec:basis}
In the following, we will introduce states with relatively
small $\Delta K$ at the expense of a nonzero but also small energy
uncertainty $\Delta H$.  
As for coherent states in the harmonic oscillator~\cite{Glauber63}, we will
minimize the product of the uncertainties involved. To do so, we consider the
uncertainty relation  
\bege\label{eq:ur}
\Delta H \Delta K \geq \frac12|\mean{[\op{H},\op{K}]}| \ .
\ende
It is an easy matter to show that the
equality in (\ref{eq:ur}) holds in a state $|L\rangle$ satisfying 
\bege\label{eq:cond}
(\op{H}-\mean{\op{H}}) |L\rangle = i a (\op{K}-\mean{\op{K}})|L\rangle \ ,
\ende
where $i^2=-1$, $a$ is an arbitrary real number and $\mean{\ldots} = \langle
L|\ldots|L\rangle$ with $\langle L|L\rangle = 1$. $|L\rangle$ has to be a
right eigenvector of the operator  
\bege\label{eq:La}
\op{L} = \op{H}-i a \op{K} \ .
\ende
$\op{L}$ does not commute with its adjoint. It therefore does not belong to
the class of normal operators, which contains Hermitian and unitary operators
as special cases. In general, the right eigenvectors of a nonnormal operator
cannot be used as a basis. Instead, right and left eigenvectors together form
a biorthogonal basis; see, e.g., \cite{Wilkinson65}. In our case, however, $\langle
E_i|\op{L}|E_j \rangle$ is a complex-symmetric matrix due to the fact that the
Hermitian matrices $\langle E_i|\op{H}|E_j \rangle$ and $K_{ij} = \langle
E_i|\op{K}|E_j \rangle$ can be made real. Left and right eigenvectors are
therefore identical and form separately a nonorthogonal basis.  
Furthermore, it can be shown that $\Delta H = |a|\Delta K$. In order to have
equal uncertainties we choose $a=1$ (states with $a=-1$ are identical). 
The real and imaginary parts of the complex eigenvalues $L$ have a
physical interpretation as mean
energies $\mean{\op{H}}$ and $-\mean{\op{K}}$, respectively.  
Note that the classical function $H-iK$ is a constant of motion
whose constant-level surfaces are the invariant surfaces.

We derive now the properties of the $|L\rangle$-states in the 
limit $\hbar \to 0$. 
Without loss of generality, we stipulate that $\op{K}$ and $K$
are homogeneous functions of the momenta of degree two, like $\op{H}$ and $H$. 
Starting from the expansion (see, e.g.,~\cite{Brenig92,HOSW84})  
\bege\label{eq:commutatorPoisson}
\langle p_x,p_y|i[\op{H},\op{K}]|x,y\rangle \langle x,y|p_x,p_y\rangle = \hbar
\{H,K\} + O(\hbar^2) \ ,
\ende
we will exploit only the fact that the Poisson bracket $\{H,K\}$ is everywhere
zero except at isolated critical points in position space. The following line
of reasoning is therefore also true for integrable billiards with a finite
number of magnetic flux lines or point scatterers (we have checked this for
{\v Seba}'s billiard~\cite{Seba90}). 
In the semiclassical regime, both sides of \equ~(\ref{eq:commutatorPoisson})
are very small in the region ${\cal Q}\setminus {\cal C}$ excluding the union
${\cal C}$ of neighborhoods of the critical corners, the area of which shrinks
to zero as $\hbar \to 0$.  
Manipulating \equ~(\ref{eq:cond}) and restricting the integration over an
$|L\rangle$-state to region ${\cal Q}\setminus {\cal C}$ gives 
\bege\label{eq:urr}
(\Delta H_{{\cal Q}\setminus {\cal C}})^2 + (\Delta K_{{\cal Q}\setminus {\cal
C}})^2 = \mean{i[\op{H},\op{K}]}_{{\cal Q}\setminus {\cal C}} \ ,
\ende
with $(\Delta H_{{\cal Q}\setminus {\cal C}})^2 =
\mean{(\op{H}-\mean{\op{H}})^2}_{{\cal Q}\setminus {\cal C}}$, etc (Note that
$\mean{1}_{{\cal Q}\setminus {\cal C}} < 1$). The
smallness of the l.h.s. of \equ~(\ref{eq:commutatorPoisson}) carries over to
both sides of \equ~(\ref{eq:urr}).
From this we conclude that a function $\langle x,y|L\rangle$ in region
${\cal Q}\setminus {\cal C}$ is locally either very small or can be
approximated by a joint eigenfunction of both operators $\op{H}$ and $\op{K}$. 
Such a joint eigenfunction cannot fulfill the boundary
conditions globally, otherwise the billiard would be integrable. Loosely
speaking, region ${{\cal Q}\setminus {\cal C}}$ must be divided into
subregions in each of which a joint eigenfunction (or a vanishing function)
fulfills the boundary conditions locally. Joint eigenfunctions of 
neighboring subregions match smoothly, so they must have roughly the same
number of nodal lines.
Hence, $|L\rangle$-states have features of energy states of integrable
systems:  
(i) the functions $\langle x,y|L\rangle$ are in some regions 
regular while in other regions, separated by ``caustics'', vanishing;
(ii) they can be labelled by two ``quantum numbers'' $(n_1,n_2)$;
(iii) the eigenvalues $L = \mean{\op{H}}-i\mean{\op{K}}$ are regularly
distributed in the complex plane.  
The last property holds because $\mean{\op{H}}$ (all arguments are also
valid for quantities derived from $\op{K}$) is asymptotically equal to
$\mean{\op{H}}_{{\cal Q}\setminus {\cal C}}$, which is approximately a
homogeneous function of the quantum numbers of degree two due to the
homogeneity of $\op{H}$. 
The non-relevance of $\mean{\op{H}}_{\cal C}$ can be understood with a
renormalization procedure. Going from $(n_1,n_2)$ to $(2n_1,2n_2)$, we get a
new wave function with slightly larger region ${{\cal Q}\setminus {\cal C}}$
containing four times more nodal lines, and region ${\cal C}$ being a four
times smaller copy of the old region ${\cal C}$.   
Hence, $\mean{\op{H}}_{{\cal Q}\setminus {\cal C}}$ roughly increases by a
factor of four, while $\mean{\op{H}}_{\cal C}$ stays constant. This leads to
$\mean{\op{H}} \to \mean{\op{H}}_{{\cal Q}\setminus {\cal C}}$ as more and
more renormalization steps bring us towards the semiclassical limit.  

The approximate homogeneity of $\mean{\op{H}}_{{\cal Q}\setminus
{\cal C}}$ with respect to the quantum numbers ensures that the local
transitions between different joint eigenfunctions induce only a small
uncertainty $(\Delta H_{{\cal Q}\setminus {\cal C}})^2$ of order
$\mean{\op{H}} \ll \mean{\op{H}^2}$, increasing roughly
by a factor of four under the action of the renormalization.  
The same factor is an upper bound for the increase of $(\Delta H_{\cal C})^2$,
corresponding to a $\mean{\op{H}^2}$-scaling weighted with the size of ${\cal
C}$.   
The sum $(\Delta H)^2 = (\Delta H_{{\cal Q}\setminus {\cal C}})^2+(\Delta
H_{\cal C})^2$ therefore increases by a factor of four. Hence,   
\bege\label{eq:limitbe}
(\Delta H)^2 \propto \mean{\op{H}}  \ , \;\;
(\Delta K)^2 \propto \mean{\op{K}} 
\ende
in the semiclassical limit. Consequently, 
\bege\label{eq:precise}
\lim_{\hbar\to 0}\frac{\Delta H}{\mean{\op{H}}} \to 0 \ , \;\;
\lim_{\hbar\to 0}\frac{\Delta K}{\mean{\op{K}}} \to 0 \ ,
\ende
i.e. the $|L\rangle$-states condense onto the invariant surfaces.  

\section{Time evolution}
\label{sec:time}
The time dependence of $|L\rangle$-states is non-trivial since $\op{L}$ does
not commute with the Hamiltonian; $|L(t)\rangle$ is in general not an
eigenstate of $\op{L}$ for $t>0$.  
Let us define three time scales associated with a given state $|L\rangle$ at
time $t=0$: the quantum mean period, the lifetime and the classical mean free
time  
\bege\label{eq:timescales}
\tauQ = \frac{2\pi\hbar}{\mean{\op{H}}} \ , \;\;
\tauL = \frac{2\pi\hbar}{\Delta H} \ , \;\;
\tauC = \sqrt{\frac{A}{2\mean{\op{H}}}} \ ,
\ende
where $A$ is the area of the billiard.
From \equ~(\ref{eq:limitbe}) we get $\tauQ \ll \tauL \approx \tauC$, i.e. the
state is quasi-stationary with a lifetime of order of the classical mean
free time.  

At first glance, it seems that the state fails to condense onto an invariant
surface for classical long times $t \gg \tauC$. However, the requirements for
condensation, small relative uncertainties and constant mean values of
$\op{H}$ and $\op{K}$, may persist beyond the lifetime of the state.  
Clearly, this is true for the relative uncertainty and the mean value of
$\op{H}$ since this operator commutes with the evolution operator
$\exp{(-i\op{H}t/\hbar)}$. 
Keeping in mind that $\mean{i[\op{H},\op{K}]}$ is relatively
small, we may expect that the relative uncertainty and the mean value of
$\op{K}$ change slower than the state itself. To see this, we
define the time scale associated with 
\bega\label{eq:Kt}
\mean{\op{K}}(t) & = & \langle L(t)|\op{K}|L(t)\rangle \nonumber\\
& = & \sum_{j,k=1}^\infty\langle L|E_j\rangle K_{jk}e^{i(E_j-E_k)t/\hbar}
\langle E_k|L\rangle\ . 
\enda 
in the same way as the lifetime in \equ~(\ref{eq:timescales}) as $\tauK =
2\pi\hbar/\Delta H_K$, where   
\bege\label{eq:dHK}
(\Delta H_K)^2 = \frac{\sum_{j,k}|\langle L|E_j\rangle K_{jk}(E_j-E_k)
\langle E_k|L\rangle|^2}{2\sum_{j,k}|\langle L|E_j\rangle K_{jk}
\langle E_k|L\rangle|^2} 
\ende
is the mean energy difference, i.e. mean frequency difference, of the
energy states involved, but in contrast to $\Delta H$ the states are weighted
according to their relevance for $\mean{\op{K}}(t)$.
Two extreme cases show that definition~(\ref{eq:dHK}) is reasonable:
a diagonal matrix $K_{jk}$ gives $\Delta H_K = 0$ whereas a
uniform matrix $K_{jk} = \const$ gives $\Delta H_K = \Delta H$.

The series~(\ref{eq:Kt}) can be expanded in orders of $t$, with $t$ much
smaller than $\tauK$ but with the possibility of being larger than the quantum
mean period and even 
the lifetime. Clearly, the zeroth-order term is $\mean{\op{K}}$. The next terms
are roughly of order $\mean{\op{K}}\Delta H_K t$, $\mean{\op{K}}(\Delta H_K
t)^2$, and so on. If we compare this to  
\bege\label{eq:taylor}
\mean{\op{K}}(t) = \mean{\op{K}} + \mean{i[\op{H},\op{K}]}t/\hbar + O(t^2) \ ,
\ende
we find that the order of $\tauK$ can be estimated as
$2\pi\mean{\op{K}}/\mean{i[\op{H},\op{K}]}$. 
It follows that in the semiclassical limit $\mean{\op{K}}(t)$ does not change
relative to its initial value for classically long times $t$ below $\tauK
\gg \tauC$, even though $|L(t)\rangle$ may differ strongly from the initial
state $|L\rangle$. 

Following the same line of reasoning shows that the time below which
$\mean{\op{K}^2}(t)$ relative to its initial value is constant scales
as $\mean{\op{K}^2}/\mean{i[\op{H},\op{K}^2]}$. Starting from 
\equ~(\ref{eq:cond}) the following equation can be derived 
\bege 
\mean{i[\op{H},\op{K}^2]}-2\mean{\op{K}}\mean{i[\op{H},\op{K}]} = 
2\mean{(\op{K}-\mean{\op{K}})^3} \ .
\ende
With the renormalization procedure it can be shown that the
asymptotic behavior of the r.h.s of this equation is bounded from above by
$\mean{\op{K}}^2$. Together with the already known result
$\mean{i[\op{H},\op{K}]} \propto \mean{\op{K}}$ we 
finally get the upper bound $\mean{i[\op{H},\op{K}^2]} \propto
\mean{\op{K}^2}$. Hence, the time scale which governs $\mean{\op{K}^2}(t)$
is of the same or of larger order as $\tauK$.  
From this we can conclude that for times smaller than $\tauK$, $\Delta K(t)$
remains small if compared with $\mean{\op{K}}(t)$. 
Hence, \equ~(\ref{eq:precise}) holds not only for times of order $\tauL
\approx \tauC$
but also for times much larger than $\tauC$ and well below $\tauK$. The
condensation of $|L\rangle$-states onto invariant surfaces outlives the
lifetime of the states up to classically long times.

\section{Example: the barrier billiard}
\label{sec:barrier}
We illustrate all statements for the barrier
billiard~\cite{Zwanzig83,HM90,Wiersig00}, a rectangle with width $l_x$ and
height $l_y$ and a vertical barrier of length $l_y/2$ placed on the symmetry
line $x=l_x/2$; see \fig~\ref{fig:barrier}(a).   
The billiard is not only pseudointegrable, it is also
almost-integrable~\cite{Gutkin86}, i.e. is composed of several copies of a
single integrable sub-billiard, here the rectangle shown in
\fig~\ref{fig:barrier}(b).   
The function $K = p_x^2$ is a second 
constant of motion. The general formula for the genus of the invariant
surfaces~\cite{RichensBerry81} gives~2, i.e. the surfaces have the topology of
two-handled spheres and not that of tori (single-handled spheres). 
\def\figbarrier{%
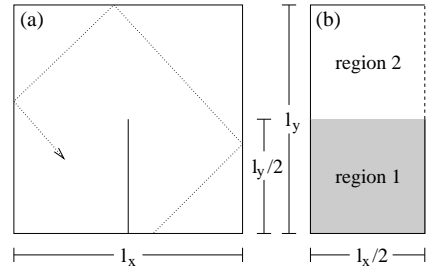
(a) Barrier billiard, rectangle with a barrier 
between the points $(x,y) = (l_x/2,0)$ and $(l_x/2,l_y/2)$. 
(b) Symmetry reduced system. The nontrivial wave functions fulfill
Dirichlet (Neumann) boundary conditions on solid (dashed) lines.}
\def\FIGbarrier{\centerline{\psfig{figure=barrier.eps,width=5.5cm,angle=0}
\vspace{0.2cm}
}}
\FIGo{fig:barrier}{\figbarrier}{\FIGbarrier}

%\subsection{Energy states}
The energy eigenfunctions are solutions of the Helm\-holtz
equation with Dirichlet boundary conditions on the polygon. The functions
are odd or even with respect to the symmetry line. The odd
ones are trivial eigenfunctions of the integrable sub-billiard. We
therefore deal only with the even ones, which fulfill mixed boundary
conditions on the symmetry reduced polygon; see
\fig~\ref{fig:barrier}(b).   
We have calculated the solutions numerically with the mode-matching method for
the parameters $\hbar =1$, $l_x = \pi\sqrt{8\pi}/3$, and $l_y =
3\sqrt{8\pi}/\pi$ as described in \app~\ref{app:energystates}. 
The statistical properties of the energy levels are found to be close to the
semi-Poisson distribution~\cite{BGS99}. The energy eigenfunctions are not
eigenfunctions of the operator $\op{K} = \op{p}^2_x$. The numerical
computation of the uncertainty $\Delta K$ is explained in
\app~\ref{app:dKE}. Figure~\ref{fig:hqes} shows that there is no trend towards
vanishing relative uncertainties. This indicates that energy states of the
barrier billiard do not condense onto the invariant surfaces.      
\def\fighqes{%
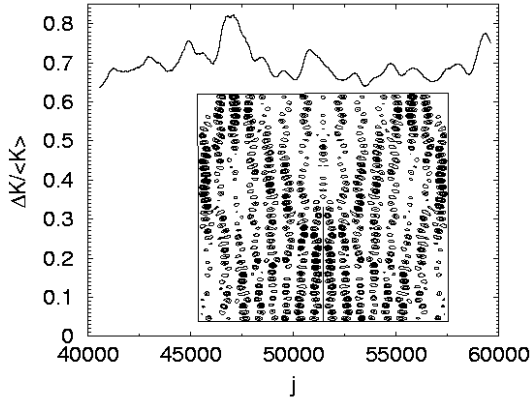
Relative uncertainty $\Delta K/\mean{\op{K}}$ of energy states $|E_j\rangle$. A
local Gaussian average with a variance of 200 is performed. 
Inset: Contour plot of the probability density associated to the 626th energy
state with $E = 646.03$, $\mean{\op{K}} = 135.05$ and $\Delta K = 144.01$.}
\def\FIGhqes{\centerline{\psfig{figure=hq.estates.eps,width=7.0cm,angle=0}
}}
\FIGo{fig:hqes}{\fighqes}{\FIGhqes}

%\subsection{$L$-states}
The nontrivial $|L\rangle$-states are also calculated with the mode-matching
method. As explained in \app~\ref{app:Lstates}, we cannot compute as
many of these states as energy states. However, even in the accessible 
low-energy regime our theoretical results on the semiclassical
behavior of these states will be well confirmed in the following. 

The regular pattern of the eigenvalues $L = \mean{\op{H}}-i\mean{\op{K}}$ can
be most clearly seen when transformed into the real action variables of the
integrable sub-billiard  
\begin{equation}\label{eq:actions}
(I_x,I_y) = \left(\frac{l_x}{2\pi}\sqrt{\mean{\op{K}}},
\frac{l_y}{\pi}\sqrt{\mean{\op{H}}-\mean{\op{K}}}\right) \ .
\end{equation}
Note that the topology of the invariant surfaces in the full system rules out
action-angle variables~\cite{RichensBerry81}.
As can be seen from \fig~\ref{fig:actions} the ``action space'' is split into
two regions A and B separated by a transition region. Away from this
transition region, and the lines $I_x = 0$ and $I_y = 0$, the eigenvalues are
approximately located on regular lattices which are given by EBK-like
quantization rules:    
\begin{equation}\label{eq:ebka}
(I_x,I_y) = \left(n_1+\frac34,n_2+1\right)\hbar 
\end{equation}
in region A and 
\begin{equation}\label{eq:ebkb}
(I_x,I_y) = \left(\frac{n_1}2+\frac12,2n_2+\frac74\right)\hbar
\end{equation}
in region B with $n_1,n_2 = 0,1,\ldots$. For example, we have $(I_x,I_y)
\approx (6.75,37)\hbar$ for the eigenfunction of type A and  $(I_x,I_y)
\approx (21,9.7)\hbar$ for the function of type B in \fig~\ref{fig:Ls}.  
Both types have relatively small uncertainties (see \app~\ref{app:dKL} for
numerical details) and look rather regular. While type-A functions cover the
billiard uniformly, apart from a localization around the critical
corner, the type-B functions are restricted to the lower ($n_1$ odd)
or upper ($n_1$ even) half of the billiard, bounded by a caustic-like curve.
\def\figactions{%
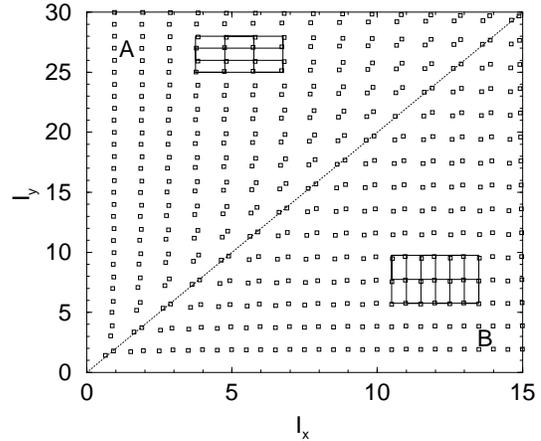
Eigenvalues $L = \mean{\op{H}}-i\mean{\op{K}}$ transformed into the
action space of the sub-billiard according to \equ~(\ref{eq:actions}).
The dotted line marks the centre of the transition region. 
The solid lines indicate parts of the EBK lattices defined in \eqs~(\ref{eq:ebka})-(\ref{eq:ebkb}).}  
\def\FIGactions{\centerline{\psfig{figure=actions.eps,width=7.0cm,angle=0}
}}
\FIGo{fig:actions}{\figactions}{\FIGactions}
\def\figLs{%
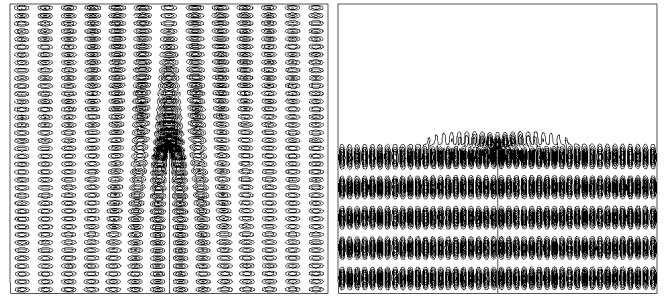
Probability density of a state $|L\rangle$ of type A (left) with eigenvalue
$656.30-i 64.93$ and $\Delta H = \Delta K = 34.88$; type B (right) with
eigenvalue $670.95-i630.52$ and $\Delta H = \Delta K = 25.14$. } 
\def\FIGLs{\centerline{
\psfig{figure=a.eps,width=4.25cm,angle=0}\vspace{0.5cm}
\psfig{figure=b.eps,width=4.25cm,angle=0}
}}
\FIGo{fig:Ls}{\figLs}{\FIGLs}

The eigenvalue pattern in \fig~\ref{fig:actions} resembles strongly that of
integrable systems with separatrices~\cite{RDWW96,WWD97,WWD98}.    
It is therefore not surprising that the statistical properties of
the mean energies of $|L\rangle$-states are similar to those of energy levels
of integrable systems. For example, \fig~\ref{fig:nnmean} confirms that the
nearest-neighbor statistics is in agreement with the Poisson distribution.  
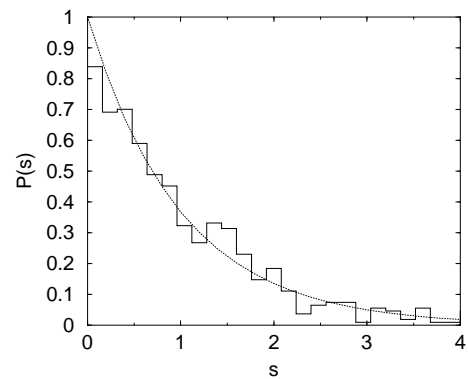
\def\fignnmean{%
Probability density $P(s)$ of the spacing $s$ between adjacent values of
the first 800 mean energies $\mean{\op{H}}$. The data is well fitted
by the Poisson distribution (dotted line).}  
\def\FIGnnmean{\centerline{\psfig{figure=nnmean.eps,width=6.0cm,angle=0}
}}
\FIGo{fig:nnmean}{\fignnmean}{\FIGnnmean}

The scaling laws~(\ref{eq:limitbe}) for the uncertainties are verified in the
following way. First note that lines of constant mean energy $\mean{\op{H}}$
are circles in the scaled action space $(2I_x/l_x,I_y/l_y)$. This plane can
therefore be conveniently parametrized by the radial coordinate
$\mean{\op{H}}$ and the polar angle $\phi =
\arctan{[(I_y/l_y)/(2I_x/l_x)]}$. The first scaling law 
holds if $(\Delta H)^2$ divided by $\mean{\op{H}}$ is a
function of $\phi$ alone. Figure~\ref{fig:hq} shows that this condition is
well satisfied.   
This confirms also the second scaling law since $\Delta K = \Delta
H$ and $\mean{\op{K}}$ scales as $\mean{\op{H}}$.
\def\fighq{%
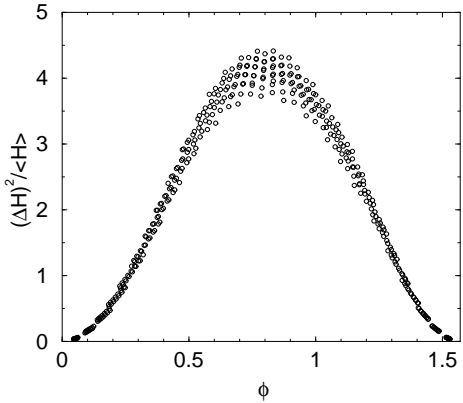
$(\Delta H)^2/\mean{\op{H}}$ vs. $\phi$ with  $400 \leq \mean{\op{H}} \leq
800$.  
The maximum lies in the transition region of action space;
cf. \fig~\ref{fig:actions}.}  
\def\FIGhq{\centerline{\psfig{figure=hq.eps,width=6.0cm,angle=0}
}}
\FIGo{fig:hq}{\fighq}{\FIGhq}

The main features of the time dependence of $|L\rangle$-states can be
observed in \figs~\ref{fig:time} and ~\ref{fig:Lst}; see \app~\ref{app:time}
for the numerical aspects. For small times $t$ around $\tauQ \ll \tauL \approx
26.7 \tauQ$, the time evolved state $|L(t)\rangle$ and the initial state
$|L\rangle$ are similar; the overlap $|\langle L(t)|L\rangle|^2$ is
close to one.  
For times $t \approx \tauL$ both states differ considerably; the
overlap is close to zero. The state has lost some of its regularity, but
$\mean{\op{K}}(t)$ and $\Delta K(t)$ stay constant up to order 
$\mean{\op{K}}$ for longer times well below $\tauK \approx 2736\tauQ$. 
Interestingly, the states of type B become more uniformly distributed in
configuration space in the course of time, cf. \figs~\ref{fig:Ls}(b), \ref{fig:Lst}(a) and (b). 
\def\figtime{%
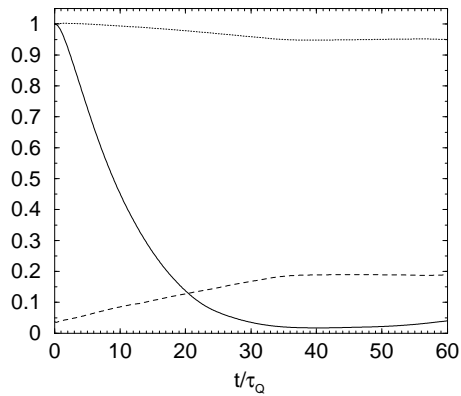
Overlap $|\langle L(t)|L\rangle|^2$ (solid),
$\mean{\op{K}}(t)/\mean{\op{K}}$ (dotted) and $\Delta
K(t)/\mean{\op{K}}(t)$ (dashed) for the state of type B with $L =
670.95-i630.52$; cf. \fig~\ref{fig:Ls}.}  
\def\FIGtime{\centerline{\psfig{figure=time.eps,width=6.0cm,angle=0}
}}
\FIGo{fig:time}{\figtime}{\FIGtime}
\def\figLst{%
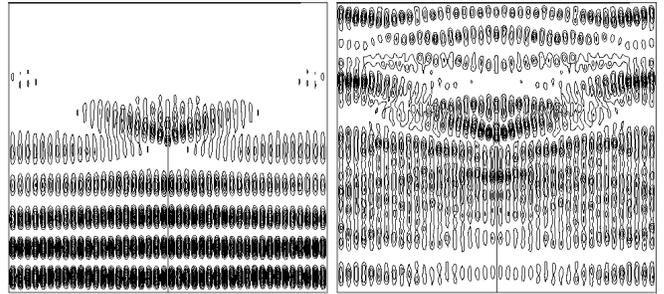
State of type B with $L = 670.95-i630.52$, cf. \fig~\ref{fig:Ls}, at time
$t=10\tauQ$ (left) and $t=\tauL\approx 26.7\tauQ$ (right).} 
\def\FIGLst{\centerline{
\psfig{figure=b1.eps,width=4.25cm,angle=0}\vspace{0.5cm}
\psfig{figure=b2.eps,width=4.25cm,angle=0}
}}
\FIGo{fig:Lst}{\figLst}{\FIGLst}

\section{conclusion}
\label{sec:con}
We have formulated the quantum-classical correspondence in rational polygonal
billiards in the following way: 
there exists a basis of quantum states with each state condensing individually
onto a classical invariant surface $H,K = \const$ up to classically long times
in the sense that the relative uncertainties $\Delta H/\mean{\op{H}}$ and
$\Delta K/\mean{\op{K}}$ vanish in the semiclassical limit.
We have presented some hints, like the analogy to the Anderson metal-insulator
transition, and numerical evidence for a particular system, the
barrier billiard, that the energy states do not show such a condensation.
We have then introduced an alternative basis of states for
which we have explicitly shown that 
(i) they are quasi-stationary;
(ii) they condense onto the invariant surfaces up to classically long
times even exceeding their lifetimes; 
(iii) in configuration space they show a regular nodal-line structure,
possibly with caustic-like boundaries; 
(iv) the eigenvalues form a regular pattern in the complex plane.  

Whether these states condense uniformly onto the invariant surfaces in
phase space is another question. At first sight, our numerical
results on the barrier billiard seem to indicate that there cannot be uniform
condensation since a fraction of the states cover only one half of the
configuration space. However, each such a state becomes more uniformly
distributed on the invariant surface for times beyond the lifetime but 
remains in the neighborhood of the surface for classically long times,
i.e. there can be uniform condensation after some transient time.

It is important to note that the condensation scenario holds for general
rational polygonal billiards and also if the
time-dependent Schr\"odinger equation is replaced by the wave equation used in
acoustics and electromagnetics, even though the individual time scales are
different.

I would like to thank M. Sieber and J. N\"ockel for discussions. 

\begin{appendix}
  \section*{Numerical computations on the barrier billiard}
\subsection{Energy states}
\label{app:energystates}
We compute the energy eigenvalues and eigenfunctions 
with the mode-matching method; see, e.g.,\cite{RichensBerry81,WSM95,ESTV96}. 
Let us first set $\hbar = 1$ and then divide the symmetry-reduced barrier
billiard in two regions as shown in \fig~\ref{fig:barrier}(b): region 1 with
$y\leq l$ and region 2 with $y > l$.  
The length of the barrier, $l$, is in our case fixed to $l_y/2$ but the
following derivations hold also for general $0<l<l_y$. 
In region 1 we expand the wave function as 
\bege\label{eq:region1}
\Phi_1 = \sum_{m=1}^\infty a_m\sin{(2m\pi x/l_x)}\sin{(g_{2m}y)}
\ende 
with 
\bege\label{eq:gE}
g^2_j(E) = E-\left(\frac{j\pi}{l_x}\right)^2 .
\ende
This kind of expansion is nontrivial since $g_j$ can be imaginary and $a_m$
complex. 
By construction, the function~(\ref{eq:region1}) fulfills the Helmholtz
equation $-\nabla^2\Phi_1=E\Phi_1$ with Dirichlet boundary condition on $x=0$, $y=0$, and $x=l_x/2$.
In region 2 we take an analog function,
\bege\label{eq:region2}
\Phi_2 = \sum_{m=1}^\infty b_m \sin{[(2m-1)\pi
x/l_x]}\sin{[g_{2m-1}(l_y-y)]} \ ,
\ende 
which satisfies Dirichlet boundary condition on $x=0$ and $y=l_y$ but Neumann
boundary condition on $x=l_x/2$.  
We stipulate that both functions match smoothly at $y=l$, i.e. for $0\leq x
\leq l_x/2$ we require 
\bege\label{eq:match1} 
\Phi_1(x,l) = \Phi_2(x,l)
\ende
and
\bege\label{eq:match2}
\frac{\partial\Phi_1}{\partial y}\Biggl|_{(x,l)} = \frac{\partial\Phi_2}{\partial y}\Biggl|_{(x,l)} .
\ende
Inserting the identity
\bege
\sin{(2n z)} = \sum_{k=1}^\infty A_{nk}\sin{[(2k-1) z]}
\ende
for $0\leq z \leq \pi/2$ with the orthogonal matrix
\bege
A_{nk} = \frac2\pi\left(\frac{\sin{[(2n-2k+1)\pi/2]}}{2n-2k+1} 
-\frac{\sin{[(2n+2k-1)\pi/2]}}{2n+2k-1}\right)
\ende
into the first matching condition~(\ref{eq:match1}) and solving for the
coefficients of $\sin{[(2n-1)\pi x/l_x]}$, $n=1,\ldots$, gives
\bege\label{eq:r1}
\sum_{m=1}^\infty a_m \sin{(g_{2m}l)} A_{mn} = b_n \sin{[g_{2n-1}(l_y-l)]} \ .
\ende
Similarly, we get from the second matching condition~(\ref{eq:match2}) 
\bege\label{eq:r2}
\sum_{m=1}^\infty a_m g_{2m}\cos{(g_{2m}l)} A_{mn} = -b_n g_{2n-1}\cos{[g_{2n-1}(l_y-l)]} .
\ende
We now rewrite the relations~(\ref{eq:r1})-(\ref{eq:r2}) by using the
definitions 
\bega
\tilde{a}_m & = & a_m g_{2m}\cos{(g_{2m}l)} ,\\
\tilde{b}_m & = & b_m g_{2m-1}\cos{[g_{2m-1}(l_y-l)]} 
\enda
and the real function
\bege\label{eq:f}
f_m(E,l) = \frac{\tan{(g_m l)}}{g_m} 
\ende
as
\bege
\sum_{m=1}^\infty \tilde{a}_m f_{2m}(E,l) A_{mn} = \tilde{b}_n f_{2n-1}(E,l_y-l) 
\ende
and
\bege\label{eq:atob}
\sum_{m=1}^\infty \tilde{a}_m A_{mn} = -\tilde{b}_n \ .
\ende
The last two equations are combined to
\bege\label{eq:lineq}
\sum_{m=1}^\infty \tilde{a}_m M_{mn} = 0 
\ende
with the real matrix
\bege\label{eq:Mmatrix}
M_{mn}(E) = [f_{2m}(E,l)+f_{2n-1}(E,l_y-l)] A_{mn} .
\ende
Equation~(\ref{eq:lineq}) has a solution provided $\det{M(E)}=0$. 
Finding the energy eigenvalues $E_i$ is therefore equivalent to finding the
zeroes of the determinant of $M$.  
We approximate $M$ by a $500\times500$ matrix which is sufficient
for calculating the first $100\, 000$ zeroes. Figure~\ref{fig:bfunc} shows
the determinant as a function of the energy.
It is convenient to search roots numerically only between two
consecutive poles. We therefore rewrite \equ~(\ref{eq:f}) using the
identity~\cite{GradRyzh65}  
\bege
-\frac{\pi}{4z}\tan{\frac{\pi z}2} = \sum_{n=1}^\infty \frac1{z^2-(2n-1)^2}
\ende
as 
\bege
f_m(E,l) = -\frac{2}{l}\sum_{n=1}^\infty \frac1{E-(\frac{m\pi}{l_x})^2-(\frac{n\pi}{2l})^2}
\ende
where the summation is only over odd $n$. From this expression the poles can be
easily read off. Note that the poles are not degenerate if $l^2/l_x^2$
and $(l_y-l)^2/l_x^2$ are irrational numbers. The interval between two given
consecutive poles is divided into 400 subintervals, each assumed to contain at
most a single change of sign of $\det{M}$. The bisection method is then
employed in order to find each zero with an accuracy $\approx 10^{-4}$ of the
mean level spacing. 
\def\figbfunc{%
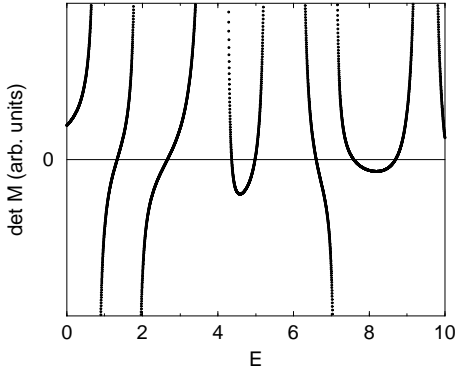
Determinant of matrix $M$ in \equ~(\protect\ref{eq:Mmatrix}).}
\def\FIGbfunc{\centerline{\psfig{figure=bfunc.eps,width=6.0cm,angle=0}
}}
\FIGo{fig:bfunc}{\figbfunc}{\FIGbfunc}

Having determined the energy levels, we get for each of them the real
quantities $\tilde{a}_m$ and $\tilde{b}_m$ from \eqs~(\ref{eq:atob}) and
(\ref{eq:lineq}) and then the wavefunction~(\ref{eq:region1}) and
(\ref{eq:region2}), which we finally normalize to unity for the full billiard.
%%%%%%%%%%%%%%%%%%%%%%%%%%%%%%%%%%%%%%%%%%%%%%%%%%%%%%%%%%%%%%%%%%%%%
\subsection{Uncertainties of energy states}
\label{app:dKE}
We here determine the uncertainty $(\Delta K)^2 =
\mean{\op{K}^2}-\mean{\op{K}}^2$ in a given $|E_j\rangle$-state.
We take $\Phi_1=\Phi_1(E_j)$, $g=g(E_j)$, and $\Phi_2=\Phi_2(E_j)$ from
\eqs~(\ref{eq:region1})-(\ref{eq:region2}) to compute  
$K_{jj} = \langle E_j|\op{K}|E_j\rangle = \mean{\op{K}}$ from   
\bega\label{eq:meanKstart}
K_{jj} & = & 
2\int_0^{l_x/2}\int_0^{l}\Phi_1^*\left(-\frac{\partial^2}{\partial x^2}\Phi_1\right) dxdy
\nonumber\\ 
& + & 2\int_0^{l_x/2}\int_l^{l_y}\Phi_2^*\left(-\frac{\partial^2}{\partial x^2}\Phi_2\right) dxdy \ .
\enda
A straightforward calculation gives
\bega\label{eq:meanKE}
K_{jj} & = & \frac{l_x}4\sum_{m=1}^\infty a^2_m h_{2m}
\left[l-\frac{\sin{(2g_{2m}l)}}{2g_{2m}}\right] \\
& + & \frac{l_x}4\sum_{m=1}^\infty b^2_m h_{2m-1}
\left[(l_y-l)-\frac{\sin{[2g_{2m-1}(l_y-l)]}}{2g_{2m-1}}\right] 
\nonumber
\enda
with $a_m = a_m(E_j)$, $b_m = b_m(E_j)$, $g_m = g_m(E_j)$, and $h_m =
(m\pi/l_x)^2$. 
Analogously, we get for $\mean{\op{K}^2} = \mean{\op{p}^4_x}$ again 
\equ~(\ref{eq:meanKE}) but with $h_m = (m\pi/l_x)^4$.

%%%%%%%%%%%%%%%%%%%%%%%%%%%%%%%%%%%%%%%%%%%%%%%%%%%%%%%%%%%%%%%%%%%%%
\subsection{$|L\rangle$-states}
\label{app:Lstates}
Eigenvalues and eigenfunctions of $\op{L}$ can be computed with the
procedure described in \app~\ref{app:energystates} if \equ~(\ref{eq:gE}) is
replaced by  
\bege\label{eq:gL}
g^2_j(L) = L-(1-i)\left(\frac{j\pi}{l_x}\right)^2 
\ende
with complex number $L$. $M$ in \equ~(\ref{eq:Mmatrix}) is then a
complex matrix. Finding the complex roots of $\det{M}$ is much more cumbersome
than finding real roots as in the case of the energy states. Because of this,
we first compute $\op{L}$ in energy-state
representation, i.e. we calculate the matrix elements $\langle
E_j|\op{L}|E_k\rangle = E_j\delta_{jk}-iK_{jk}$ with $K_{jk} = \langle
E_j|\op{K}|E_k\rangle$ given by \equ~(\ref{eq:meanKE}) if $j=k$, otherwise
\bega
K_{jk} & = & \frac{l_x}4\sum_{m=1}^\infty a^\dagger_m a_m h_{2m}
\left[\frac{\sin{(g^-_{2m}l)}}{g^-_{2m}}-
\frac{\sin{(g^+_{2m}l)}}{g^+_{2m}}
\right] \nonumber\\
\label{eq:Kij}
& + & \frac{l_x}4\sum_{m=1}^\infty b^\dagger_m b_m h_{2m-1}\times\\
& & \left[\frac{\sin{[g^-_{2m-1}(l_y-l)]}}{g^-_{2m-1}}-
\frac{\sin{[g^+_{2m-1}(l_y-l)]}}{g^+_{2m-1}}
\right]\nonumber
\enda
with $a^\dagger_m = a^*_m(E_j)$, $a_m = a_m(E_k)$, $b^\dagger_m = b^*_m(E_j)$, $b_m =
b_m(E_k)$, $g^+_m = g^*_m(E_j)+g_m(E_k)$, $g^-_m = g^*_m(E_j)-g_m(E_k)$, and
$h_m = (m\pi/l_x)^2$. 
A LAPACK routine is used to diagonalize the complex matrix $\langle
E_j|\op{L}|E_k\rangle$ with
$j,k\leq 1000$, giving a rough approximation to the first 800 eigenvalues
of $\op{L}$. We use them as initial guesses for Newton's method in order to
find reliable approximations to the eigenvalues.  
Finally, we compute for each determined eigenvalue the complex quantities
$\tilde{a}_m$ and $\tilde{b}_m$ from \eqs~(\ref{eq:atob}) and (\ref{eq:lineq})
and then the normalized wavefunction~(\ref{eq:region1}) and
(\ref{eq:region2}).    
%%%%%%%%%%%%%%%%%%%%%%%%%%%%%%%%%%%%%%%%%%%%%%%%%%%%%%%%%%%%%%%%%%%%%
\subsection{Uncertainties of $|L\rangle$-states}
\label{app:dKL}
The main advantage of using the mode-matching method for the
$|L\rangle$-states is that we get the uncertainties $\Delta K$ and $\Delta
H$ as accurately as for the energy states in \app~\ref{app:dKE}. We take
$\Phi_1=\Phi_1(L_j)$, $\Phi_2=\Phi_2(L_j)$, and $g = g(L_j)$ from
\eqs~(\ref{eq:region1}), (\ref{eq:region2}), and~(\ref{eq:gL}) to compute
firstly $\mean{\op{K}}$ starting from \equ~(\ref{eq:meanKstart}).
A straightforward calculation shows that $\mean{\op{K}}$ is given by
the r.h.s. of \equ~(\ref{eq:Kij}) with $a^\dagger_m = a^*_m(L_j)$, $a_m =
a_m(L_j)$, $b^\dagger_m = b^*_m(L_j)$, $b_m = b_m(L_j)$, 
$g^+_m = g^*_m(L_j)+g_m(L_j)$, $g^-_m = g^*_m(L_j)-g_m(L_j)$, and $h_m =
(m\pi/l_x)^2$. 
Similarly, we get for $\mean{\op{K}^2} = \mean{\op{p}^4_x}$ the same equation
but with $h_m = (m\pi/l_x)^4$; for $\mean{\op{H}}$ we
use $h_m = g^2_m(L_j)+(m\pi/l_x)^2$ and for $\mean{\op{H}^2}$ we use $h_m =
|g^2_m(L_j)+(m\pi/l_x)^2|^2$.

%%%%%%%%%%%%%%%%%%%%%%%%%%%%%%%%%%%%%%%%%%%%%%%%%%%%%%%%%%%%%%%%%%%%%
\subsection{Time dependence of $|L\rangle$-states}
\label{app:time}
We here compute $\langle x,y|L(t)\rangle$, $\langle L(t)|L\rangle$,
$\mean{\op{K}}(t)$ and $\mean{\op{K}^2}(t)$. These expressions can be written
in terms of energy states as   
\bega
\langle x,y|L(t)\rangle & = &\sum_{j=1}^\infty e^{-iE_jt}\langle E_j|L\rangle \langle
x,y|E_j\rangle\ , \\
\langle L(t)|L\rangle & = &\sum_{j=1}^\infty e^{iE_jt}|\langle E_j|L\rangle|^2 \ , 
\enda
and as in \equ~(\ref{eq:Kt}). The first 1400 energy states are incorporated in
these sums. 
As in the previous sections it can be shown that $\langle
E_j|L\rangle$ is given by the r.h.s. of 
\equ~(\ref{eq:Kij}) with $a^\dagger_m = a^*_m(E_j)$, $a_m = a_m(L)$,
$b^\dagger_m = b^*_m(E_j)$, $b_m = b_m(L)$, $g^+_m = g^*_m(E_j)+g_m(L)$,
$g^-_m = g^*_m(E_j)-g_m(L)$, and $h_m = 1$. 
$\langle E_j|\op{K}^2|E_k\rangle$ is also given by the r.h.s. of
\equ~(\ref{eq:Kij}) if $j\neq k$ with $a^\dagger_m = a^*_m(E_j)$, $a_m =
a_m(E_k)$, $b^\dagger_m = b^*_m(E_j)$, $b_m = b_m(E_k)$, $g^+_m =
g^*_m(E_j)+g_m(E_k)$, $g^-_m = g^*_m(E_j)-g_m(E_k)$, and $h_m =
(m\pi/l_x)^4$. For $j=k$ one has to use \equ~(\ref{eq:meanKE}).

\end{appendix}

%\clearpage
\bibliographystyle{prsty}
\bibliography{}

\end{multicols}

\end{document}